\documentclass[12pt]{article}

\usepackage{amssymb}
\usepackage{bm}
\newcommand\be{\begin{equation}}
\newcommand\ee{\end{equation}}
\newcommand\no{\nonumber\\}
\newcommand\beqa{\begin{eqnarray}}
\newcommand\eeqa{\end{eqnarray}}
\newcommand\bea{\begin{eqnarray*}}
\newcommand\eea{\end{eqnarray*}}

\newcommand\tet{\theta}
\newcommand\la{\lambda}
\newcommand\lb{\label}
\newcommand\m{\mu}
\newcommand\n{\nu}
\newcommand\al{\alpha}

\newcommand\de{\delta}
\newcommand\s{\sigma}
\newcommand\e{\epsilon}

\newcommand\A{{\cal A}}
\newcommand\B{{\cal B}}
\newcommand\F{{\cal F}}
\newcommand\La{{\cal L}}
\newcommand\Ha{{\cal H}}

\newcommand\cO{{\cal O}}

\newcommand\fr{\frac}
\newcommand\del{\partial}
\newcommand\h{\hat}
\newcommand\ti{\tilde}

\newcommand\lp{\left(}
\newcommand\rp{\right)}
\newcommand{\eij}{\epsilon_{ij}}

\begin{document}
\begin{center}

{\LARGE Spin Hall Effect in Noncommutative Coordinates  }

\vspace{15mm}

\"{O}mer F. Dayi$^{a,b}$
and
Mahmut Elbistan$^{a,}$\footnote{E-mail addresses:
dayi@gursey.gov.tr and elbistan@itu.edu.tr.}

\vspace{5mm}

\noindent{\it $a)$Physics Department, Faculty of Science and
Letters, Istanbul Technical University, TR-34469 Maslak--Istanbul, Turkey,}

and

\noindent {\it $b)$ Feza G\"{u}rsey Institute, P.O. Box 6,
TR--34684, \c{C}engelk\"{o}y--Istanbul, Turkey. }

\vspace{1cm}

{\bf Abstract}

\end{center}

{\small
A semiclassical constrained Hamiltonian system which was established to
study dynamical systems of matrix valued non--Abelian gauge fields
is employed  to formulate spin Hall effect in noncommuting coordinates
at the first order in the  constant noncommutativity parameter $\theta .$
The method is first illustrated by studying
the Hall effect on the noncommutative plane
in a gauge independent fashion.
Then,
the Drude model type and the Hall effect type
formulations of  spin Hall effect are considered in noncommuting coordinates and
$\theta$ deformed spin Hall conductivities which they provide are acquired.
It is shown that by adjusting $\theta$
different formulations of spin Hall conductivity are accomplished.
Hence, the noncommutative theory
can be envisaged  as  an effective theory which unifies different approaches
to  similar physical phenomena.  }

\vspace{1cm}

\section{Introduction}

Deformation quantization  which is also known as  Weyl--Wigner--Groenewold--Moyal
method of quantization\cite{wwgm}, although developed  as an alternative
to operator quantization of quantum mechanics became one of the main
approaches of incorporating noncommutative coordinates into quantum mechanics.
In this scheme one introduces
star product of coordinates in terms of the constant, antisymmetric  noncommutativity parameter
$\tet_{\mu\nu} $  and then ordinary multiplication is replaced with  star product
in  energy eigenvalue problems\cite{mez}.
In the quantum phase space given by $(\h x_\mu ,\h p_\mu )$ this scheme  is equivalent
to shift the  coordinates in the related Hamiltonian by
\be
\lb{bsh}
\h x_\mu\rightarrow \h x_\mu-\frac{1}{\hbar}\tet_{\mu \nu}  \h p_\nu ,
\ee
as far as the first order terms in $\tet$ are retained.
 The $\tet$ deformed Hamiltonian   can also   be employed
to define the related path integrals (see \cite{ao} and the references given therein).
When  gauge fields are present in the original theory
this procedure is very sensitive to the
explicit realization of the gauge field.
Moreover, it is not suitable to envisage how spin dependent particles would behave in
noncommutative coordinates.
Therefore, it is desirable to establish a systematic method of formulating dynamics in noncommutative
coordinates which does not refer  to a particular   gauge choice
as well as embraces spin dependent systems.
Recently spin dependent dynamics
was studied as a semiclassical constrained Hamiltonian system\cite{om}.
We will show that this constrained system is suitable to formulate
either spinless or spin dependent dynamical systems  in noncommutating
coordinates. Moreover, this approach is gauge independent
as far as the related canonical Hamiltonian is chosen appropriately.
Some spinless dynamical systems in noncommuting coordinates
have already been considered as  constrained Hamiltonian systems\cite{der}.
However, the basic achievement reported here is to offer  a
 systematic method of studying
matrix
valued observables in
 noncommuting coordinates.

When the noncommutativity of coordinates is introduced
without an apparent physical motivation, there is a subtle issue:
How the noncommutativity of coordinates in dynamical systems should be  interpreted?
Obviously,  it can be taken literally claiming that
its effects have not been observed yet due to the smallness of the noncommutativity parameter $\tet$.
However, we prefer another interpretation which sounds more
realistic:
The dynamical system in noncommuting coordinates
effectively
links different manifestations of the same theory. It  either relates
non--interacting theory to the interacting ones or connects different
kinds of interactions which explain similar phenomena\cite{ao},\cite{sc}.
We would like to apply this point of view to obtain an effective theory of
spin Hall effect which has been studied recently
 in terms of different methods\cite{mnz}--\cite{smeh} but  an interrelation is missing
(for a review see \cite{erh}).
We will consider two simple formulations of  spin Hall effect.
The first is an extension of the Drude model\cite{chu} and the other
is obtained as a generalization of the Hall effect employing the Rashba spin--orbit coupling\cite{om}.
Studying these formulations in noncommutative space leads to  deformed spin Hall
conductivities which reproduce
 the non--deformed ones
when the noncommutativity is turned off by setting $\tet =0.$ In this respect it differs from
the
deformed spin Hall conductivity obtained in \cite{ar} which vanishes in the $\tet =0$  limit.
In \cite{ar} two dimensional harmonic oscillator in noncommutative coordinates was
considered adopting
a similar interpretation of the noncommutativity.

We will introduce noncommutativity of  coordinates within
the semiclassical constrained Hamiltonian system which is  briefly reviewed in  Section 2.
To illustrate the method
we first apply it
to ``ordinary'' Hall effect in Section 3 in a gauge invariant fashion. We also show that when
gauge fields are  incorporated into the  Hamiltonian,
manifestations of the noncommutativity depend on the  choice of  gauge.
In Section 4 we study the spin Hall effect formulations  of \cite{chu} and \cite{om}
in noncommutating coordinates and obtain  $\tet$ deformed spin Hall conductivities which they
provide.
Then, we
discuss how to  obtain  the spin Hall conductivities  furnished with some different
approaches
from the deformed spin Hall conductivity
by fixing the noncommutativity parameter $\tet.$
We only deal with the deformations which are at the first order in $\theta .$

\section{The semiclassical approach}

We would like to recall  the main ingredients of
the semiclassical approach established in \cite{om} for studying spin dependent dynamical
systems.
It is based on an extension of the deformation quantization\cite{wwgm}:
Let us deal with matrix observables
which may depend on $\hbar ,$ although they are functions of the  classical phase space variables
$(\pi_\mu , x_\mu ).$   In terms of the star product
$$
\star   =
\exp \left[
\frac{i\hbar }{2} \left(
\frac{\overleftarrow{\del}}{\del x^\m}
\frac{\overrightarrow{\del}}{\del \pi_\m}
-\frac{\overleftarrow{\del}}{\del \pi^\m}
\frac{\overrightarrow{\del}}{\del x_\m}
\right) \right] , \label{sct}
$$
one introduces the Moyal bracket of
the matrix observables  $K_{ab}(\pi ,x)$ and $N_{ab}(\pi ,x)$  as
\be
\left( [K(\pi,x ) , N(\pi,x ) ]_\star \right)_{ab}  =  K_{ac}(\pi,x ) \star N_{cb}(\pi,x )
- N_{ac}(\pi,x ) \star K_{cb}(\pi,x ) . \lb{MM}
\ee
The semiclassical approximation is introduced by  the bracket
\be
 \{K(\pi,x ) , N(\pi,x ) \}_C \equiv
\frac{-i}{\hbar}[K,N]+
\frac{1}{2} \{ K(\pi,x ) , N(\pi,x ) \}
-\frac{1}{2} \{ N(\pi,x ) , K(\pi,x ) \} , \lb{SCB}
\ee
obtained from
the Moyal bracket (\ref{MM}) by retaining the lowest two terms in $\hbar$.
The first term on the right hand side  is the ordinary commutator of matrices and
the others are  Poisson brackets defined as
$$
 \{ K(\pi,x ) , N(\pi,x ) \} \equiv
  \frac{\partial K}{\partial x^\n} \frac{\partial N}{\partial \pi_\n}
-\frac{\partial K}{\partial \pi_\n}\frac{\partial N}{\partial x^\n} . \lb{ocl}
$$
It is a semiclassical approximation because in (\ref{SCB}), where the observables
$K$ and $N$ may depend on $\hbar ,$ only the two lowest order terms in $\hbar$ are detained.

Semiclassical  dynamical equations are given by
employing the semiclassical bracket (\ref{SCB}) to define
the time evolution of a semiclassical observable $K(\pi,x)$ as
$$
{\dot K } (\pi,x) = \{K(\pi,x ) , H(\pi,x ) \}_C ,\lb{evo}
$$
in terms of the
semiclassical Hamiltonian  $H(\pi ,x).$
As usual the dot
over the observables indicates the time
derivative.

\subsection{A semiclassical constrained Hamiltonian system}

Semiclassical Hamiltonian dynamics is
 developed emulating the rules of
the ordinary Hamiltonian
formulation by replacing Poisson brackets
with the semiclassical brackets (\ref{SCB}).
It is  a systematic method of introducing and studying  noncommutativity of
phase space variables.  Let us consider  the gauge fields
$\A_\al , \B_\al ;$ $\al =1,\cdots, n ;$
  which are in general  $N\times N$ matrices
and the  underlying first order matrix Hamiltonian
\be
\La  = {\dot r}^\al \lp \fr{1}{2}I p_\al +\rho \A_\al (r,p)  \rp \no
-{\dot p}^\al \lp \fr{1}{2}I r_\al -\xi \B_\al (r,p) \rp -\Ha_0(r,p). \lb{oL}
\ee
$\rho , \xi   $
are the coupling constants and
$I$ denotes the unit matrix.
Canonical momenta  are defined as usual:
$$
\pi_r^\al =\frac{\del \La}{\del \dot r_\al} ,\ \pi_p^\al =\frac{\del \La}{\del \dot p_\al} .
$$
Being a first order  Hamiltonian  (\ref{oL}) leads to the primary constraints
\beqa
\psi^{1\al} & \equiv & (\pi_r^\al -\fr{1}{2}p^\al )I -\rho \A^\al ,\lb{pcs1}\\
\psi^{2\al} & \equiv & (\pi_p^\al +\fr{1}{2}r^\al )I -\xi \B^\al .\lb{pcs2}
\eeqa
In terms of  the  Lagrange multipliers  $\la^\al_z;$   $z=1,2,$
and the
canonical Hamiltonian $\Ha_0 $  one  introduces the extended Hamiltonian
\be
\lb{eh}
\Ha_e =\Ha_0 +\la^\al_z\psi^z_\al .
\ee
To employ
the semiclassical approach we set
$\pi^\mu=(\pi^\al_p,\pi^\al_r)$ and $x_\mu =(p_\al ,r_\al ).$
Now, one can observe that the constraints obey the semiclassical brackets
\beqa
\{\psi^1_\al ,\psi^1_\beta \}_C & = & \rho F_{\al \beta} ,\nonumber \\
\{\psi^2_\al ,\psi^2_\beta \}_C & = & \xi G_{\al \beta},\nonumber \\
\{\psi^1_\al ,\psi^2_\beta \}_C & = & -\delta_{\al \beta} -M_{\al \beta} , \nonumber
\eeqa
where $\delta_{\al\beta}$ is  the Kronecker delta and the field strengths are defined by
\begin{eqnarray*}
F_{\al \beta} & = &
\fr{\del \A_\beta}{\del r^\al}
-\fr{\del \A_\al}{\del r^\beta}
-\fr{i\rho }{\hbar} [\A_\al ,\A_\beta ] ,\\
G_{\al\beta}  & = &
\fr{\del \B_\beta}{\del p^\al}
-\fr{\del \B_\al}{\del p^\beta}
-\fr{i\xi }{\hbar} [\B_\al ,\B_\beta ] ,\\
M_{\al \beta} & = & \xi \fr{\del \B_\beta}{\del r^\al}
-\rho \fr{\del \A_\al}{\del p^\beta}
-\fr{i\xi\rho}{\hbar} [\A_\al ,\B_\beta ]  .\lb{M}
\end{eqnarray*}
For consistency the constraints (\ref{pcs1}) and (\ref{pcs2})
should be constant in time:
\be
\lb{sco}
\{ \psi^z_\al , \Ha_e \}_C\approx 0,
\ee
where $\approx$ denotes
that the equality is satisfied on the  hypersurface where  constraints   vanish.
The conditions (\ref{sco}) are solved by fixing the Lagrange multipliers as
\be
\lb{lm}
\la^\al_z =-\{\psi^{z^\prime}_\beta ,H_0\}_C C_{z z^\prime }^{-1 \al \beta} ,
\ee
where
we utilized the definitions
$$
\lb{cm}
C^{zz^\prime}_{\al \beta}=
\{\psi^z_\al , \psi^{z^\prime}_\beta\}_C	;\
C^{zz^{\prime \prime}}_{\al \gamma }
C_{z\prime z^{\prime\prime}}^{-1\gamma \beta}=\delta_\al^\beta \delta_{z^\prime}^z .
$$

To impose vanishing of the second class constraints (\ref{pcs1}), (\ref{pcs2})
effectively, one introduces the semiclassical Dirac bracket
\begin{equation}
\label{sdb}
\{K,N\}_{CD} \equiv \{K,N\}_C -\{K,\psi^z\}_C C^{-1}_{zz^\prime}\{\psi^{z^\prime},N\}_C  .
\end{equation}
Extending  the usual Dirac procedure we replace
the semiclassical bracket of observables (\ref{SCB})  with
the semiclassical Dirac bracket (\ref{sdb}) in dynamical equations.
Hence, the phase space variables  satisfy the relations
\beqa
\{r^\al,r^\beta \}_{CD}  & = &  C_{11}^{-1\al\beta} =
\xi  G^{\al\beta  }
-\xi (MG)^{\al\beta  }
  + \xi (MG)^{\beta \al}  -\rho\xi^2(GFG)^{\al\beta}+\cdots
,\lb{rr}\\
\{p^\al,p^\beta \}_{CD}  & = &  C_{22}^{-1\al\beta} =
\rho  F^{\al\beta  }
- \rho (MF)^{\al\beta  }
  +\rho (MF)^{\beta \al}  -\rho^2\xi(FGF)^{\al\beta} +\cdots ,\lb{yy}\\
\{r^\al,p^\beta \}_{CD}  & = &  C_{12}^{-1\al\beta} =
\delta^{\al \beta }+ M^{ \beta \al} - \rho\xi (GF)^{\al\beta}
-(MM)^{\al\beta}+\cdots  .\lb{ry}
\eeqa
Observe that
$r_\al$ and $p_\al$  should be considered as coordinates and the corresponding
momenta, respectively, after effectively eliminating the constraints
(\ref{pcs1}), (\ref{pcs2}).

Consistent with the  usual procedure
$$
\lb{eqm}
{\dot \cO }(r,p)  =  \{\cO (r,p) ,H_e\}_C
$$
defines the
equation of motion of the observable $\cO (r,p)$.
$ H_e $ is defined in (\ref{eh}) with the Lagrange multipliers given as in (\ref{lm}).
Therefore,  one can find that
\beqa
{\dot r}^\al &  = &
 \xi \lp \fr{\del H_0}{\del r^\beta} -\fr{i\rho}{\hbar} [ \A_\beta  ,H_0]
\rp
\left(  G^{\al\beta  }
-(MG)^{\al\beta  }
  + (MG)^{\beta \al}  -\rho\xi(GFG)^{\al\beta}+\cdots
\right) \no
&&  +\lp\fr{\del H_0}{\del p^\beta} -\fr{i\xi}{\hbar} [\B_\beta ,H_0] \rp
 \lp \delta^{\al \beta }+ M^{ \beta \al} - \rho\xi (GF)^{\al\beta}
-(MM)^{\al\beta}+\cdots \rp , \lb{gss1}
\\
{\dot p}^\al & = &
\lp\fr{\del H_0}{\del r^\beta} -\fr{i\rho}{\hbar} [\A_\beta  ,H_0]
\rp \lp-\delta^{\al \beta } - M^{ \al \beta }
+ \rho \xi (FG)^{ \al\beta} +(MM)^{\al\beta} +\cdots \rp
\no
&& +\rho \lp \fr{\del H_0}{\del p^\beta} -\fr{i\xi}{\hbar} [\B_\beta ,H_0] \rp
\lp   F^{\al\beta  }
- (MF)^{\al\beta  }
  +(MF)^{\beta \al}  -\rho\xi(FGF)^{\al\beta} +\cdots
\rp  ,\lb{gss2}
\eeqa
are the equations of motion of the phase space variables.
Although this formalism is valid in any  dimensions, in the following  we will consider physical systems which are
effectively 2--dimensional.

\section{Hall Effect in Noncommutative Coordinates}

Electrons moving in a thin slab in the presence of uniform  external
magnetic field perpendicular to the plane  will experience the Lorentz force.
Hence they will be pushed on a side of the slab
 producing a potential difference between the sides. This is known as Hall  effect.
If one applies an external electric field balancing the potential difference,
electrons will move without deflection\cite{gir}. This approach gives a simple
formalism of deriving the Hall conductivity\cite{om}.
We would like to study this system in noncommuting coordinates.
Although we
still use the terminology of Section 2, as far as spinless systems are considered semiclassical
brackets coincide with their classical counterparts.

Let us deal with the  Hamiltonian
\be
\lb{hal0}
H_0= \fr{1}{2m} \left(p_1^2 +p_2^2\right)+ V (\bm r ) ,
\ee
in terms of  the scalar potential
\be
\lb{esp}
V (\bm r )=-eE_ir_i
\ee
yielding the uniform external electric
field  $E_i;$ $i=1,2.$ We consider an electron moving on the $r_1r_2$ plane but
we choose the coupling constant to be  $\rho =e/c $ and
 the  $\A_i$ field such that there is a uniform magnetic field $\bm B=B\h r_3$ perpendicular to the
plane of motion:
\be
\lb{fsf}
F_{ij}=B\e_{ij}.
\ee

By inspecting the relation (\ref{rr}) it is evident that
to introduce  noncommutating coordinates we need to choose
\be
\lb{gij}
G_{ij}=\e_{ij}
\ee
and identify the related coupling constant with
 the noncommutativity parameter:  $\xi =\theta .$
The gauge field yielding (\ref{gij}) can be defined for example by
$\B_i=-\eij p_j/2 $, though  its specific form is not needed.
Employing (\ref{fsf})--(\ref{gij})
 in (\ref{rr})--(\ref{ry}) and retaining the terms at the
 first order in $\tet$ and $eB/c ,$ one can show that
\beqa
\{r_i,r_j \}_{CD}  & = &   \theta \eij , \lb{ch1}  \\
\{p_i,p_j \}_{CD}  & = &     \fr{eB}{c} \eij ,\lb{ch2} \\
\{r_i,p_j \}_{CD}  & = &   \left( 1+ \fr{e B \tet}{c}  \right)\delta_{ij} .\lb{ch3}
\eeqa
In this formalism $p_i$ act as kinematic momenta (\ref{ch2})  and due to this fact the semiclassical bracket
(\ref{ch3}) possesses a term depending on the noncommutativity parameter $\tet ,$
in contrary to the formalisms where canonical momenta are adopted (see (\ref{ca1})--(\ref{ca3})).

Keeping the terms
at the first order in $\tet$ and $eB/c ,$
(\ref{gss1}) and (\ref{gss2}) yield the following
equations of motion
of the phase space variables
\beqa
\dot r_i & = & -e\tet \eij E_j +\left(1+ \fr{e B \tet}{c} \right)\fr{ p_i}{m} ,\lb{dhal11} \\
\dot p_i & = &  \left(1+ \fr{e B \tet}{c} \right) eE_i +\fr{eB}{mc}\eij p_j, \lb{dhal12}
\eeqa
Expressing $p_i$ in terms of the
velocity
$v_i\equiv \dot r_i ,$
the force acting on  the electron  follows from (\ref{dhal11}) and (\ref{dhal12}) as
$$
\F_i = m \ddot{r}_i = \left(1+ 2\fr{eB  \tet}{c} \right)
e E_i+\fr{eB}{mc}\eij \left(v_j+e\tet \epsilon_{jk}E_k\right) . \lb{dfh}\\
$$
For not being deflected the force acting on the electron should vanish:
$$
\lb{fe}
\F_i =0,
$$
which can be solved for the velocity as
\be
\lb{vif}
v_i= \fr{c}{B}\left(1+ \fr{e B\tet }{c} \right)\eij E_j.
\ee
We would like to associate the above formulation with a system of
electrons. To this aim let us introduce the density of electrons $\kappa$
and define the electric current  as
\be
\lb{cure}
j_i=e\kappa v_i.
\ee
Employing the velocity obtained  demanding that
the electrons move without deflection   (\ref{vif}) in (\ref{cure})
 yields the electric current
$$
j_i= -\sigma_H (\tet ) \eij E_j ,
$$
where the deformed Hall conductivity is
\be
\lb{gis}
\sigma_H (\tet ) =-\left(1+ \fr{e B \tet}{c} \right)\frac{ec\kappa}{B}.
\ee

 (\ref{gis}) is gauge invariant in the sense that it does not depend on how
the vector potential $\A_i$
 is realized explicitly.
(\ref{gis}) depends on the field strength (\ref{fsf}).
Deformation of Hall conductivity in noncommuting coordinates
has already been addressed in \cite{ao}. However,
there is a discrepancy between  the deformed Hall conductivities
reported there and the one obtained here (\ref{gis}).
The reason resides in the fact that
the deformation of Hall conductivity obtained in \cite{ao} by the custom method
(\ref{bsh}), depends on how the gauge field is realized. In \cite{ao} vector potential
was in symmetric gauge. We would like to show that within our  approach,
by choosing the
Hamiltonian suitable to the custom method of deformation one obtains the same factors of
deformation in symmetric gauge but other choices lead to a non--deformed Hall
conductivity.
Deal with the Hamiltonian
$$
H_0=\frac{1}{2m}\left(  p_i-(e/c) A_i \right)^2+V(\bm r),
$$
where the scalar potential
is still as in (\ref{esp}) and the
noncommutativity furnished through
(\ref{gij}). There is no other gauge field.
Hence, $p_i$ behave as canonical momenta and
the deformed canonical relations are now given by
\beqa
\{r_i,r_j \}_{CD}  & = &   \theta \eij , \lb{ca1}  \\
\{p_i,p_j \}_{CD}  & = &    0 ,\lb{ca2} \\
\{r_i,p_j \}_{CD}  & = &   \delta_{ij} .\lb{ca3}
\eeqa

In  the symmetric  gauge
$$
 A_i=- \fr{B}{2}\eij r_j,
$$
from (\ref{rr})--(\ref{ry})
 we obtain the equations of motion
\bea
\dot r_i & = & -e\tet \eij E_j +\left(1+ \fr{eB\tet}{2c} \right)
\left(\fr{ p_i}{m}-\fr{e}{mc}A_i \right), \\
\dot p_i & = &  eE_i +\fr{eB}{2c}\eij \left(\fr{ p_j}{m}-\fr{e}{mc}A_j \right).
\eea
Now, following the procedure given above one can solve the condition
$m\ddot{r}_i=0$ for the velocity $\dot r_i$ and plug into the current
(\ref{cure}). This yields the following deformed Hall conductivity
$$
\sigma_H^S (\tet ) =-\left(1- \fr{eB\tet }{4c} \right)\frac{ec\kappa}{B} ,
$$
which is the result obtained in \cite{ao}
up to a rescaling of the noncommutativity parameter with $\hbar .$
The rescaling is needed  due
to the fact that the noncommutativity parameter of \cite{ao}
is obtained like (\ref{bsh}) with another $\hbar^{-1}$ factor.
This result is gauge dependent. Indeed,
if one adopts other gauge choices like
$ A_i=(-Br_2,0)$ or $A_i=(0,Br_1),$
following the same procedure
one can show that
there will be no $\tet$ dependent contribution to Hall conductivity
though  coordinates are noncommuting.

We would like to elucidate the idea of interpreting a non interacting dynamical system in
noncommutating coordinates as an effectively interacting one\cite{ao},\cite{sc}
considering the
fractional quantum Hall effect whose conductivity is
$$
\s_H^F=\nu \frac{e^2}{h} ,
$$
where $\nu =1/3,2/3,1/5,\cdots .$
The idea is to fix the noncommutativity parameter $\tet$ as
$$
\tet_F  =-\fr{\nu}{\kappa h}-\frac{c}{eB},
$$
so that (\ref{gis}) yields
$$
\sigma_H (\tet )|_{\tet =\tet_F} =\s_H^F.
$$
Hence, the fractional quantum  Hall effect where electrons are interacting
can be obtained from the Hall effect in noncommutative space
which is  a non--interacting theory.
Thus, one avoids to deal with a complicated interacting theory
favoring  an effective theory which is simpler.

\section{Spin Hall Effect in Noncommutative Coordinates}

To understand spin Hall effect,
which basically occurs due to spin currents produced by spin--orbit coupling terms in the
presence of electric field, diverse models were developed\cite{mnz}--\cite{smeh}
(for a complete list see  \cite{erh}).
However,
there are two simple
semiclassical formulations which are suitable to investigate spin Hall effect
in noncommuting coordinates: $i)$ The extension of the Drude model given
in \cite{chu}. $ii)$ Generalization of  the  Hall effect formulation
proposed in \cite{om}.

\subsection{Deformation of the Drude model type formulation}

The extension of the Drude model discussed in \cite{chu} can  be obtained within the
semiclassical approach of Section 2.  The appropriate
gauge field is
\be
\lb{AGF}
\A_a=\fr{\e_{a b c}\s_b}{4m c^2} \fr{\del V}{\del r_c},
\ee
where $a=1,2,3;$ $\s_a$ are the Pauli spin matrices and  we set the coupling constant to be
$\rho=-\hbar .$
Without ignoring the $1/c^4$ terms, the related field strength becomes
\be
\lb{AFF}
F_{ab}=\fr{\e_{b c d}\s_c}{4m c^2} \fr{\del^2 V}{\del r_a\del r_d}
-\fr{\e_{a c d }\s_c}{4m c^2} \fr{\del^2 V}{\del r_b \del r_d}
-\fr{\e_{a b c  }}{8m^2 c^4} \bm\s \cdot \bm \nabla V \fr{\del V}{\del r_c} .
\ee

We deal with an electron moving on the noncommutative $r_1r_2$ plane, so that
like the previous section we consider the Hamiltonian (\ref{hal0}) and set
$\xi=\tet$ and $G_{ij}=\eij .$ Hence,
ignoring the terms at the order of $\hbar^2$ one can show from (\ref{rr})--(\ref{ry}) that
 the phase space variables satisfy
\beqa
\{r_i,r_j \}_{CD}  & = &  \theta  \eij , \lb{sh1}  \\
\{p_i,p_j \}_{CD}  & = &  -\hbar F_{i j}   ,\lb{sh2} \\
\{r_i,p_j \}_{CD}  & = &
\delta_{ij}+\hbar \tet \epsilon_{ik} F_{k j}.
\lb{sh3}
\eeqa
As before
$p_i$ are kinematic momenta thus the $\tet$ dependent term appears in  (\ref{sh3}).

The equations of motion following from (\ref{gss1}) and (\ref{gss2}) are
\beqa
\dot r_i & = &
\fr{ p_i}{m} +
\fr{\hbar \tet }{m}\epsilon_{ij} F_{ jk}p_k
+\tet \eij \fr{\del V}{\del r_j} ,\lb{ce1} \\
\dot p_i & = &  -\fr{\del V}{\del r_i}
-\hbar \tet  F_{ij}\epsilon_{jk}\fr{\del V}{\del r_k}
 -\fr{\hbar}{m} F_{ij}p_j . \lb{ce2}
\eeqa
In the spirit of
 the Drude model by adding the drag force
$-p_i / \tau $ where  $\tau$ is the relaxation  time
and retaining the terms linear in  the velocity $v_i$, (\ref{ce1}), (\ref{ce2})
yield  the total force
\beqa
\F_i   =  m \ddot{r}_i-\fr{p_i}{\tau} & =&
-\fr{\del V}{\del r_i}
-\hbar \tet  F_{ij}\epsilon_{jk}\fr{\del V}{\del r_k}
-\hbar F_{ij}v_j +\tet \eij v_k \fr{\del^2 V}{\del r_j \del r_k}
 -\fr{m}{\tau}
 v_i \no
&& + \fr{m\hbar \tet}{\tau}
\epsilon_{ij}F_{jk}v_k
+ \fr{m\tet}{\tau} \epsilon_{ij}\fr{\del V}{\del r_j}.   \lb{dfsh}
\eeqa
The  $\hbar^2$ terms are ignored and we used (\ref{ce1}) to express the momentum $p_i$
in terms of the velocity $v_i$ as
\be
\lb{pi}
\fr{p_i}{m}= v_i - \tet \epsilon_{ij}\fr{\del V}{\del r_j}
-\hbar \tet \epsilon_{ij}F_{jk}v_k.
\ee
In the absence of the drag force and for  $\tet=0$
the  force (\ref{dfsh}) has already been derived in \cite{shen} employing the
gauge field (\ref{AGF}) and
 also in \cite{om} within  another approach.

According to \cite{chu} let us deal with
\be
V= V_l +V_e
\ee
where $V_l$ is the potential collecting effects of crystal and
$$
V_e=-e\bm E\cdot \bm r
$$
is the external scalar potential.
Moreover, in the force (\ref{dfsh}) we replace potential terms
 with their volume averages.
Considering cubic lattice the following average over the volume
is given in terms of the constant $A$ as
$$
\left< \fr{\del^2 V}{\del r_a \del r_b}\right> =-eA\de_{ab}.
$$
Because of the external electric field $\bm E$ we also have
$$
\left< \fr{\del V}{\del r_a }\right> =-eE_a.
$$
Hence, average value of the field strength (\ref{AFF})  is
\be
\lb{ft}
\left< F_{ij}\right> =-\left( \fr{eA}{2mc^2}\s_3 +\fr{e^2}{8m^2c^4}\bm \s \cdot \bm E  E_3 \right)\eij .
\ee
In \cite{kra} it is claimed that the constant $A$ should vanish due to electrical neutrality
of conductors. On the other hand in \cite{chu1}  it is argued that the original formulation
\cite{chu} makes use of the localized charges leading to an estimate for $A$ which is
nonvanishing (see also \cite{st}).
Nevertheless, as far as the average (\ref{ft}) is  nonvanishing the extension
of Drude model is successful
 though the estimated value of spin Hall conductivity may differ.

One demands that electron moves with a constant velocity,
so that  the total force acting on the
electron should vanish:
\be
\lb{f0}
 \F_i =0.
\ee
Plugging (\ref{ft}) into (\ref{f0}) yields
\be
\lb{st}
eE_i -\fr{mv_i}{\tau} -e\tet\eij \left( Av_j -\fr{mE_j}{\tau}  \right)
+\left( \fr{e\hbar A}{2mc^2}\s_3 +\fr{e^2\hbar }{8m^2c^4}\bm \s \cdot \bm E  E_3 \right)
\left( e\tet E_i +\eij v_j+\fr{m\tet v_i}{\tau}\right) =0.
\ee
The steady--state solution can be obtained treating the $\left< F_{ij}\right>$ dependent
contribution in (\ref{st}) as perturbation. Hence let
$$
v_i=v_i^0+v_i^I
$$
where the lowest order solution is
\be
\lb{v0}
v_i^0=\fr{e\tau}{m} E_i .
\ee
The first order solution in the perturbation can be shown to be
\beqa
v_i^I & = &
\fr{2e\tau\tet}{m}\left( \fr{e\hbar A}{2mc^2}\s_3 +\fr{e^2\hbar }{8m^2c^4}\bm \s \cdot \bm E  E_3 \right) E_i
\no
&&
-\left[ e\tet \left(-1 +\fr{\tau^2 e A}{m} \right) + \fr{\tau^2 e }{m}
\left( \fr{e\hbar A}{2mc^2}\s_3 +\fr{e^2\hbar }{8m^2c^4}\bm \s \cdot \bm E  E_3 \right)
 \right] \eij E_j. \lb{vi}
\eeqa
Let us introduce the density matrix
\be
\lb{dm}
N=\fr{1}{2}n\left(1+\bm\xi \cdot \bm \s \right)
\ee
in terms of
the spin polarization vector  $\bm \xi$  whose magnitude is
$$
\xi =\fr{n^\uparrow-n^\downarrow}{n^\uparrow +n^\downarrow},
$$
where $n^\uparrow$ and $n^\downarrow$ denote concentration of states with spins
along the $\h \xi$ and $-\h \xi $ directions and $n=n^\uparrow +n^\downarrow .$
We choose the spin polarization to point in the third direction:
$$\bm \xi=\xi \h r_3 .$$
Adopting the definition of the  current given in \cite{chu} as
\be
\lb{cuc}
 j_i \equiv e{\rm Tr}\left( N v_i \right)
\ee
and making use of the steady--state solution given by (\ref{v0})--(\ref{vi}), we obtain
$$
\lb{cuc1}
j_i =\s_C(\tet)E_i-\s^D_{SH} (\tet )  \xi \eij  E_j.
$$
where the $\tet$ deformed conductivity is
$$
\s_C(\tet)=
\left(1+
 \fr{e\hbar \tet A}{mc^2}+\fr{e^2\hbar\tet E_3^2 }{4m^2c^4}\right) \fr{ne^2\tau}{m}
$$
and the $\tet$ deformed spin Hall conductivity is
\be
\lb{esh}
\s_{SH}^{D} (\tet )=
- \fr{n\hbar\tau^2 e^3 A   }{2m^3c^2}
- \fr{n\hbar\tau^2 e^4 E_3^2   }{8m^4c^4}
+ \frac{ne^2\tet}{\xi}\left(1+\fr{e\tau^2A}{m}\right) .
\ee
Indeed, when one sets  $\tet=0$ and ignore $1/c^4$ terms the formalism of \cite{chu} follows.

\subsection{Deformation of the Hall Effect type formulation}

Another simple method of deriving spin Hall conductivity
was developed in \cite{om} generalizing the formulation of Hall effect presented
in Section 3.
In this approach one introduces the non--Abelian
gauge field
\be
\lb{ars}
\A_i=\e_{ij}\s_j  ,
\ee
which is consistent with the  Rashba spin--orbit coupling term\cite{ras}.
It leads to
\be
\lb{fra}
F_{ij}=-\frac{i\rho}{\hbar}[\A_i,\A_j]=\frac{2\rho}{\hbar}
\sigma_3 \e_{ij}.
\ee
The main ingredients of the dynamical equations (\ref{rr})--(\ref{gss2}) are the field
strengths.
Explicit forms of gauge fields do not matter as far as they
lead to the same field strengths and commute with the related
canonical Hamiltonian.
We will deal with the Hamiltonian (\ref{hal0}) which is a scalar, so that
instead of the gauge field (\ref{ars}) we may equivalently choose
\be
\lb{adr}
\A_i=\s_i  ,
\ee
yielding the same field strength  (\ref{fra}) up to a minus sign.
The gauge field (\ref{adr}) is consistent with the
Dresselhaus spin--orbit coupling term\cite{dre}.
Obviously, we may also let the both gauge fields (\ref{ars}) and (\ref{adr})
be present by different coupling constants.

As before noncommutativity of the $r_1r_2\equiv xy$ plane is furnished by
letting the coupling constant be $\xi=\tet $ and
\be
\lb{gra}
G_{ij}=\eij .
\ee
Moreover, as it is announced the canonical Hamiltonian is given by
(\ref{hal0}) with the scalar potential (\ref{esp}).

By plugging (\ref{fra}), (\ref{gra}) into
(\ref{rr})--(\ref{ry}) and keeping only the first order terms in $\tet$ and $\rho^2 $
one obtains the relations
\begin{eqnarray*}
\{r_i,r_j \}_{CD}  & = &  \theta  \eij , \lb{sh1d}  \\
\{p_i,p_j \}_{CD}  & = &  \fr{2\rho^2}{\hbar}\s_3 \eij   ,\lb{sh2d} \\
\{r_i,p_j \}_{CD}  & = & \left( 1+\fr{2\rho^2\tet }{\hbar}\s_3 \right) \delta_{ij}  .\lb{sh3d}
\end{eqnarray*}
As it is typical to our formalism $p_i$ act as kinematic momenta.

At the same order in $\tet$ and $\rho ,$
making use of (\ref{fra}),(\ref{gra}) in (\ref{gss1}),(\ref{gss2})
the equations of motion of the phase space variables  are obtained as
\begin{eqnarray*}
\dot r_i & = & \left( 1+\fr{2\rho^2\tet }{\hbar}\s_3 \right)\fr{ p_i}{m} -e\tet \eij E_j, \\
\dot p_i & = & \left( 1+\fr{2\rho^2\tet }{\hbar}\s_3 \right) eE_i +\fr{2\rho^2}{\hbar m} \s_3 \eij p_j .
\end{eqnarray*}
In terms of the velocity $  v_i \equiv \dot  r_i ,$
we can write the momentum in the first order in $\tet$ as
$$
\fr{p_i}{m}=\left( 1-\fr{2\rho^2\tet }{\hbar}\s_3 \right) v_i+e\tet \eij E_j.
$$
Hence, at the first order in $\tet$ and $\rho^2$
the force acting on the particle is
$$
\lb{raf1}
\F_i=m\ddot r_i  =eE_i  + \frac{2\rho^2}{\hbar} \sigma_3\e_{ij}  v_j
+\frac{2e\rho^2\tet }{\hbar}\s_3E_i .
$$
Imitating the formulation of the Hall effect  we demand that
\be
\label{0F}
 \F_i =0
\ee
in order to have a motion without deflection.
The condition (\ref{0F}) can be  solved for  the velocity as
\beqa
v_i^\uparrow  &=&  \left( \frac{e\hbar}{2\rho^2} +e\tet \right) \eij E_j,  \lb{v1u}  \\
 v_i^\downarrow & = & -\left( \frac{e\hbar}{2\rho^2} -e\tet \right) \eij E_j,  \lb{v2d}
\eeqa
The arrows
$\uparrow$ and $\downarrow$ correspond to the positive
and negative eigenvalue of the Pauli spin matrix $\s_3.$ It is natural to define
the spin current   as
\be
\lb{cur}
j_i^z=\frac{\hbar}{2}
\left(n^\uparrow v_i^\uparrow - n^\downarrow  v_i^\downarrow \right),
\ee
where $n^\uparrow$ and $n^\downarrow$ denote concentration of states with spins
along the $\h r_3\equiv \h z$ and $-\h z$ directions.  Employing (\ref{v1u}),(\ref{v2d}) in
(\ref{cur}) yields
$$
\lb{cur1}
j_i^z =-\s_{SH}(\tet)   \eij E_j,
$$
where now, the $\tet$ deformed  spin Hall conductivity is given by
\be
\lb{shc}
\s_{SH}(\tet)=\frac{-e\hbar^2 n}{4\rho^2} -\frac{1}{2}e\hbar n\xi\tet .
\ee
Let $n=n^\uparrow +n^\downarrow$  be given as the concentration of states occupying the
lower energy state of the Rashba Hamiltonian\cite{ras}
times a constant $l :$
$$
\lb{n}
n= \frac{\rho^2 l}{\pi \hbar^2}.
$$
Denote that the original
Rashba coupling constant
$\al$  is related to $\rho$ as
 $ \rho =-\al m/ \hbar $\cite{ras}.
Then,
the $\tet$  deformed spin Hall conductivity can be written as
\be
\lb{ish}
\s_{SH}(\tet)=  -\frac{el}{4\pi} -\frac{e\hbar \ti\tet }{2}
\ee
where we defined
$$
\ti \tet \equiv \left( n^\uparrow -n^\downarrow   \right) \tet  .
$$
Observe that,  the $\tet=0$ limit of  (\ref{ish})
 agrees with the universal behavior obtained in \cite{sin} for  $l=1/2 .$

\subsection{Deformed Spin Hall Conductivity}

The difference in the  $\xi$ dependence of  (\ref{esh}) and (\ref{ish})
is due to the fact that in the former the density matrix (\ref{dm})
is employed to define the current (\ref{cuc}),
 but in the latter  we avoided it in the definition of the spin current (\ref{cur}).
Nevertheless,
both of the formalisms which we presented here lead to  deformed spin Hall conductivities
which can be written symbolically as
$$
\lb{gdshc}
\s_{S}(\tet )=\Sigma_0 +\tet \Sigma_1 ,
$$
which yields spin Hall conductivity when the noncommutativity is switched off
by setting $\tet=0.$
In \cite{ar} another $\tet$ deformed spin Hall conductivity was  delivered
 which   vanishes in the limit $\tet =0.$

For concreteness in the following discussions we will concentrate on  (\ref{ish}).
In the spirit of interpreting the noncommutativity as a link between similar physical phenomena
 $\tet$  can be
fixed to obtain  other formulations of spin Hall effect.
We will illustrate this point of view by considering
spin Hall conductivities obtained
by inclusion of impurities, the
Rashba type spin--orbit couplings with higher order momenta
and the quantum spin Hall effect.

When impurity effects included into the
 Rashba Hamiltonian which is linear in momenta,
the universal behavior of spin Hall conductivity\cite{sin}
is swept out\cite{vsh}.
This can be obtained by
fixing the noncommutativity parameter in  (\ref{ish}) as
$$
\ti \tet_0= -\frac{l}{2\pi\hbar }
$$
 yielding
$$
\s_{SH}(\tet)|_{\ti \tet =\ti \tet_0} =0.
$$

On the other hand when one deals with  the Rashba  type Hamiltonian with  higher order momenta
$$
H=\epsilon_k-\fr{1}{2} b_i (\bm k)  \s_i +V(\bm r) ,
$$
one finds a non--vanishing spin Hall conductivity\cite{smeh}.
Here $\bm k$ is the kinematic momentum and $\epsilon_k$
is the energy dispersion in the absence of spin--orbit coupling.
 By defining
$b_1+ib_2\equiv b_0(k)\exp (iN\tet )$ and
$$
\ti N =\fr{d \ln |b_0|}{d \ln k },\ 1+\zeta = \fr{d \ln v}{d \ln k },
$$
in terms of the velocity  $v$,
the spin Hall conductivity
$$
\s_{SH}^{HR}=-\fr{eN}{4\pi  }\left(\frac{N^2-1}{N^2+1}\right)
\lp\ti N -\zeta -2\rp
$$
results.
This can be achieved from (\ref{ish}) as
$$
\s_{SH}(\tet)|_{\ti \tet =\ti \tet_{HR}} =\s_{SH}^{HR},
$$
by setting $l=N$ and
fixing the noncommutativity parameter as
$$
\ti \tet_{HR} =
\fr{N}{2\pi \hbar}\left[-1+ \left(\frac{N^2-1}{N^2+1}\right)\lp\ti N -\zeta -2\rp\right] .
$$

Quantization of the spin Hall conductance in units of $e/2\pi$ was predicted in \cite{bz}.
Hence the quantized spin Hall conductivity can be written as
$$
\s_{SH}^{Q}= -\fr{e}{2 \pi}\mu
$$
where $\mu$ is a number depending on the physical system considered. This can be obtained from
(\ref{ish})  by fixing the noncommutativity parameter as
$$
\s_{SH}(\tet)|_{\ti \tet =\ti \tet_{Q}} =\s_{SH}^{Q},
$$
where
$$
\ti\tet_Q=\frac{1}{2\pi \hbar}\lp -l +2\mu \rp.
$$

Therefore, the spin Hall effect in noncommutative coordinates can be considered as the
master formulation such that fixing the noncommutativity parameter $\tet$
yields different manifestations  of the same physical  phenomenon.


\begin{thebibliography}{99}

\bibitem{wwgm}H. Weyl, Z. Phys. 46 (1927) 1;\\
E. Wigner. Phys. Rev. 40 (1932) 749;\\
H.J. Groenewold, Physica 12 (1946) 405;\\
J. Moyal, Proc. Cambridge Philos. Soc. 45 (1949) 99.

\bibitem{mez}L. Mezincescu,``Star operation in quantum mechanics", hep-th/0007046.

\bibitem{ao}\"{O}. F. Dayi and A. Jellal, J. Math. Phys. 43 (2002) 4592 [Erratum: 45 (2004) 827].

\bibitem{om}\"{O}. F. Dayi,  J. Phys. A: Math. Theor. 41 (2008) 315204.

\bibitem{der}A.A. Deriglazov, Phys. Lett. B 530 (2002) 235.

\bibitem{sc}F.G. Scholtz, B. Chakraborty, S. Gangopadhyay and A. G. Hazra, Phys. Rev. D 71 2005 085005;\\
F.G. Scholtz, B. Chakraborty, S. Gangopadhyay and J. Govaerts, J. Phys. A: Math. Gen. 38 (2005) 9849.

\bibitem{mnz}S. Murakami, N. Nagaosa and S-C. Zhang, Science, 301 (2003) 1348.

\bibitem{sin}J. Sinova, D. Culcer, Q. Niu, N.A. Sinitsyn, T. Jungwirth and
A.H. MacDonald, Phys. Rev. Lett. 92 (2004) 126603.

\bibitem{vsh}J-I. Inoue, G.E.W. Bauer and L.W. Molenkamp,
Phys. Rev. B 70 (2004) 041303(R);\\
R.Raimondi and P. Schwab, Phys. Rev. B 71 (2005) 033311;\\
O.V. Dimitrova, Phys. Rev. B 71 (2005) 245327.

\bibitem{smeh}A.V. Shytov, E.G. Mishchenko, H-A. Engel and B.I. Halperin, Phys. Rev. B (2006) 075316.

\bibitem{erh}H-A. Engel, E.I. Rashba and B.I. Halperin, in: H. Kronm\"uller and S. Parkin (Eds.),  Handbook of Magnetism and Advanced Magnetic Materials,. John Wiley \& Sons Ltd, Chichester, UK, pp 2858-2877 (2007), cond-mat/0603306.

\bibitem{chu}E.M. Chudnovsky, Phys. Rev. Lett. 99 (2007) 206601.

\bibitem{ar}A. Jellal and R. Hou\c{c}a, ``Noncommutative Description of Quantum Spin Hall Effect",
 hep-th/0611301.

\bibitem{gir}S.M. Girvin, in: Topological Aspects of Low Dimensional Systems, 
ed. A. Comtet, T. Jolicoeur, S. Ouvry, F. David (Springer-Verlag, Berlin and Les Editions de Physique, Les Ulis, 2000), cond-mat/9907002.

\bibitem{shen}S-Q. Shen,  Phys. Rev. Lett. 95 (2005) 187203.

\bibitem{kra}V. Ya. Kravchenko, Phys. Rev. Lett. 100 (2008) 199703.

\bibitem{chu1}E.M. Chudnovsky, Phys. Rev. Lett. 100 (2008) 199704.

\bibitem{st}M. Schulz and S. Trimper, Phys. Lett. A 372 (2008) 5905.

\bibitem{ras}Y.A. Bychkov and E.I. Rashba, J. Phys. C 17 (1984) 6039.

\bibitem{dre}D. Dresselhaus, Phys. Rev. 100 (1955) 580.

\bibitem{bz}B.A. Bernevig and S-C. Zhang, Phys. Rev. Lett. 96 (2006) 106802.



\end{thebibliography}
\end{document}